\documentclass{wqcd03}                 

\usepackage{txfonts}                   

\def\lQ{\Lambda_{\rm QCD}}

\newcommand{\be}{\begin{equation}}
\newcommand{\ee}{\end{equation}}
\newcommand{\bea}{\begin{eqnarray}}
\newcommand{\eea}{\end{eqnarray}}

\def\als{\alpha_{\rm s}}
\def\siml{{\ \lower-1.2pt\vbox{\hbox{\rlap{$<$}\lower6pt\vbox{\hbox{$\sim$}}}}\ }} 
\def\simg{{\ \lower-1.2pt\vbox{\hbox{\rlap{$>$}\lower6pt\vbox{\hbox{$\sim$}}}}\ }}

\def\vbfD{{\
    \lower-8pt\vbox{\hbox{\rlap{$\!\leftrightarrow$}\lower8pt\vbox{\hbox{$\!\bf D$}}}}\ }}

\confname{QCD@Work 2003 - International Workshop on QCD, Conversano, Italy, 
14--18 June 2003}

\title{Potential models, sum rules, color singlet model versus NREFT
in  Heavy Quarkonium}

\author{Nora~Brambilla\addressmark{a}
}

\address[a]{Dipartimento di Fisica and INFN, Via Celoria 16, 20133 Milano, Italy}

\begin{document}

\begin{abstract}
I briefly review how nonrelativistic effective field theories 
solve old puzzles and  open problems in heavy quarkonium physics.
\end{abstract}

\maketitle


\section{Historical recollection}
 Today quarkonium is a key issue in 
most of the accelerator experiments (see \cite{qwg,thom}) where millions
and in some cases billions of heavy quarkonium states are being produced.
Historically quarkonium has been one of the most important playgrounds
for our understanding of the strong interactions. 

In perspective, the seventies were dominated by the $J/\psi$ discovery, around 3095 MeV 
with a lifetime about 1000 times longer 
than that of other particles of comparable mass. This was known in the 
particle physics community as the  November revolution. During the one year after the discovery 
more than seven hundreds papers were written related to $J/\psi$.
Most of the papers dealt with quark and antiquark in a colour singlet state 
bound by a phenomenological potential. Since the structure of the energy levels
of charmonium and the soonly after discovered bottomonium is somehow intermediate 
bewteen a Coulomb and an harmonic oscillator structure and due to the
asymptotic freedom idea \cite{app}, the basic form of the static potential 
is taken as  a superposition of a Coulomb and a linear potential:
\begin{equation}
V_0 = -{4\over 3} {\als \over r} + \sigma r
\end{equation}
where $\als$, the strong coupling constant at some 
scale, and $\sigma$, the string tension,   have to be fit on the data.
Such static singlet potential, together with similar phenomenological 
spin-dependent corrections, when used inside a Schr\"odinger equation
gives an overall successfull description of heavy quarkonium spectra and decays \cite{pot,fen,thom}.
However, already in these pioneering 
papers 
the problem arised of the 
relation of these potential models to QCD and to the QCD parameters.
Starting with the work of Wilson, an effort   \cite{wilsonpot,rev} 
was devoted  to relate the singlet potential, static and relativistic 
corrections, to QCD average values of Wilson loop and field strength insertions 
into the Wilson loop, which are objects easy to evaluate 
in lattice QCD \cite{bali}. 
However, it turned out that these results for the potentials
in terms of Wilson loops were inconsistent with the one loop calculation 
of the potentials in perturbation theory \cite{rev}.

The color singlet  potential  model encountered soon concrete problems
also  in the calculation of 
inclusive annihilation decays rates 
of heavy quarkonium states into light hadrons
(hadronic)  and photons and lepton pairs (electromagnetic). 
It was assumed that the decay rate of the quarkonium state 
factored into a short distance part $f$, calculated 
in perturbative QCD as the annihilation rate 
of the heavy quark and antiquark, 
and a long distance nonperturbative part given in terms of 
 the quarkonium wave function  (or its derivatives) evaluated at the origin:
\begin{equation}
\Gamma = f(\als(m)) \, \cdot \, \vert \psi(0)\vert^2 .
\label{decpot}
\end{equation}
Explicit calculations at next to leading order in $\als$ 
in perturbation theory for $S$ and for $P$ wave decays into photons supported  
 the factorization assumption which could not, however, be proved on general 
grounds for higher orders of perturbation theory. Indeed, 
in the case 
of $P$ wave decays into light hadrons, it turned out that at order $\als^3$ 
 the factorization was spoiled by logarithmic infrared divergences. 
The same problem appeared in relativistic corrections to the annihilation 
decays of $S$ wave states \cite{pqcd}. Therefore,  potential models
were unable to supply  at higher order an infrared finite prediction 
for the inclusive decays.

The eighties witnessed the success of the sum rules. It was a distinctive prediction 
of the sum rules that as a result of QCD   the $\eta_c$ mass has to be 
located around 3 GeV  and not at  the  much smaller value of 2.83 GeV claimed at that time by 
the experiments \cite{sumrul}. The later discovery of the genuine $\eta_c$ state with mass 
2.98 Gev at Stanford  was a great success of QCD and gave  motivations for further work on the 
sum rules. Sum rules work in terms of Wilson operator  expansion. The Green function of
the quark-antiquark pair  injected in the vacuum  is expanded in terms of 
perturbative coefficients and local nonperturbative objects like the gluon condensate and the quark 
condensate. Physically as far as the $Q\bar{Q}$ distance $r$ is held
smaller than the confinement scale $\lQ^{-1}$ the binding bewteen the quark is 
perturbative and  the propagation of the heavy pair 
is  taking place in nonperturbative ``external'' vacuum fields. In such situation the interaction 
with the nonperturbative (at the scale $\lQ$)  vacuum gluonic fields can be expanded in multipoles
 and the leading contribution of the vacuum fields 
to the quarkonium energy levels is proportional to the gluon condensate 
$ \langle 0 \vert F^a_{\mu\nu}(0) F^a_{\mu\nu}(0)\vert 0\rangle$ \cite{sumrul2}.
Such condensate corrections are incompatible with  the genuine potential described in the previous 
paragraph. Precisely, the effects of the nonperturbative fluctuations
of the  gluonic field cannot be expressed in terms of a singlet $Q\bar{Q}$  interaction potential 
 and one should consider states  that contain both the $Q\bar{Q}$ pair and a gluonic excitation, 
including the case in which both of them are in a color octet \cite{sumrul2}.
In brief, sum rules results for heavy quarkonia seemed to be in strong 
contradiction with the potential model picture.

The nineties were dominated by the data on  prompt production of charmonium at Fermilab.
The first hadron collider measurements of inclusive charmonium production  at CERN and 
from the CDF collaboration at Fermilab could not separate charmonium produced in hard scattering reactions 
from charmonium produced in weak decays of $B$ mesons. Thus the comparison with theory was uncertain.
A rigorous test of the colour singlet production model (i.e. the assumption that 
the two quark were produced in a color singlet state) became possible with the CDF data on direct 
charmonium production  where the contributions from the $B$-decays had been removed using microvertex-detection. 
With these data it became clear that the colour-singlet model failed dramatically when confronted 
with the experimental results\cite{prod}.
 
This historical excursion ends with several puzzles and open problems
related to the existence or nonexistence of a quark antiquark singlet potential,
and the validity or not of a color singlet picture.
Today the stage is taken by the many  experiments accumulating
high statistic data samples at quarkonium resonances and with several production 
mechanisms.  It becomes therefore  even more relevant to clarify the open theoretical problems and to 
supply a clean and under control QCD picture of these systems.
In the next section we will show how nonrelativistic effective field theories
provide the solution.

\section{Non Relativistic Effective Field Theories for Heavy Quarkonium}

The reason for which the EFT approach is so successfull for heavy quarkonium 
is the fact  that heavy quarkonium, being a non-relativistic bound state, 
is characterized by a hierarchy of energy scales $m$, $mv$ and $mv^2$, with  $m$  the heavy-quark 
mass and $v\ll 1$ the relative heavy-quark velocity.
A hierarchy of EFTs may be constructed by systematically integrating out 
modes associated to the energy scales not relevant for the quarkonium system.
Such integration  is made  in a matching procedure that 
enforces the complete equivalence between QCD and the EFT at a given 
order of the expansion in $v$ and $\als$.

 Integrating out degrees of freedom 
of energy $m$, which for heavy quarks can be done perturbatively, leads to  
non-relativistic QCD (NRQCD)\cite{nrqcd1,nrqcd2}. This EFT still contains the lower 
energy scales as dynamical degrees of freedom. In the last years, the problem 
of integrating out the remaining dynamical scales of NRQCD has been addressed 
by several groups and has reached  a good level of 
understanding (a  list of references can be found in \cite{reveft}). 
The EFT obtained by subsequent matchings from QCD, where only the
lightest degrees of freedom of energy $mv^2$ are left dynamical, 
is potential NRQCD, pNRQCD \cite{Mont,long}. This EFT is close to a
quantum-mechanical description of the bound system and, therefore, as simple. 
It has been systematically explored in the dynamical
regime $\lQ \siml mv^2$ in \cite{long,logs,logss} and in the regime $mv^2 \ll \lQ \siml mv$ 
in \cite{long,M12,sw}. The quantity $\lQ$ stands for the generic scale of 
non-perturbative physics.

Inside the  EFT,  the power counting in the small quantity $v$ enables to select 
the operators that contribute to physical quantities up to a definite 
order in $v$. The EFT approach  makes it possible, in the case of several
observables, to achieve a rigorous factorization between 
 the high-energy dynamics encoded into matching coefficients 
calculable in perturbation theory and the non-perturbative QCD dynamics 
encoded  into few well-defined nonperturbative contributions  
to be fitted on the data or calculated on the lattice.
Thus, several model independent QCD predictions become possible. 
I will detail these in the following section for NRQCD and pNRQCD.

\section{Non Relativistic QCD}

NRQCD is the EFT obtained by integrating out 
the hard scale $m$. The mass $m$ being larger than the scale 
of non-perturbative physics, $\lQ$, the matching to NRQCD can be done 
order by order in $\als$. Hence, the NRQCD Lagrangian can be written 
as a sum of terms like $ f_n \, O^{(d_n)}_n/m^{d_n-4}$, ordered in powers of $\als$ 
and $v$. More specifically, the Wilson coefficients 
$f_n$ are series in $\als $ and encode the ultraviolet physics that  
has been integrated out from QCD.
 The operators $O^{(d_n)}_n$ of dimension $d_n$
describe the low-energy dynamics and are counted in powers of $v$.
Heavy quarkonium annihilations are controlled by the imaginary part 
of the NRQCD Hamiltonian, i.e. the 
imaginary part of the Wilson coefficients of the 4-fermion operators 
($O^{(d_n)}_n = \psi^\dagger K_n  \chi \chi^\dagger K^\prime_n \psi$) 
in the NRQCD Lagrangian.
The wave function of the quarkonium state  
is given by a series of terms in which the leading one
is the  quark antiquark in a color singlet state and the first 
correction, suppressed in $v$, comes from quark-antiquark in an 
octet state with a gluon:
\begin{equation}
\vert H\rangle = (\vert Q \bar{Q}_1 \rangle + \vert Q \bar{Q}_8 g \rangle + \dots )
\otimes \vert nljs \rangle .
\end{equation}
To calculate physical quantities like spectra and decays the operators have to be 
evaluated over the wave functions and the power counting of the two combines to give 
the order in $v$ of the calculation. It is then clear that in the case in which 
the octet operators are enhanced with respect to the singlet, the octet part 
can be as relevant as the singlet one. The EFT contains naturally  octet 
contributions.

\subsection{Spectrum}
The NRQCD Lagrangian is well suited for lattice evaluation. The quark propagators 
are the nonrelativistic ones and since we have integrated out the scale of the mass, 
the lattice step used in the simulation may be a factor $1/v$ bigger. Lattice evaluation 
of heavy systems like bottomonium become thus feasible. The latest results for the spectra 
(quenched and unquenched) are given e.g. in \cite{davies}. The radial splittings are accurate 
up to order $O(\als v^2)$ while fine and hyperfine splittings are accurate only up to $O(\als)$,
due to the fact that only tree level matching coefficients have been used. A calculation 
of the NRQCD matching coefficients in the lattice regularization scheme is still missing
and would be relevant to improve the precision of the lattice data.

\subsection{Decays}
NRQCD gives a factorization formula for heavy
quarkonium (H) inclusive decay widths into light hadrons (LH)  \cite{nrqcd2}
\begin{eqnarray}
\!\!& &\Gamma({\rm H}\to{\rm LH}) = 
\sum_n {2 \, {\rm Im} \, f_n \over m^{d_n - 4}}
\; \langle {\rm H} | \psi^\dagger K_n  \chi \chi^\dagger K^\prime_n \psi |{\rm H} \rangle.
\label{fac1}
\end{eqnarray}
Similar formulas hold for the electromagnetic decays.
The 4-fermion operators are classified with respect to their rotational 
and spin symmetry (e.g. $O(^{2S+1}S_J)$, $O(^{2S+1}P_J)$, ...) and of their 
colour content (octet, $O_8$, and singlet, $O_1$, operators).
Singlet operator expectation values may be easily related to the square 
of the quarkonium wave functions (or derivatives of it) at the origin. 
Octet contributions remain as  nonperturbative matrix elements 
of operators over the quarkonium wave functions.
According to the power counting of NRQCD, the 
octet  contribution $\langle h | O_8 (^1S_0) | h \rangle$ to $P$-wave decays
is as relevant as the singlet contribution \cite{nrqcd2}.
This octet contribution reabsorbs the dependence 
on the infrared cut-off  of the Wilson coefficients 
solving the problem that arised in the color singlet potential model.
Systematic
improvements are possible, either by calculating higher-order corrections 
in the coupling constant or by adding higher-order operators. If one goes 
on in the expansion in $v$, that seems to be necessary for charmonium,
the numbers of the involved nonperturbative matrix elements of octet operators
over quarkonium states
increases in such a  way that limits the prediction power.

Besides this, precise theoretical predictions are also hampered by uncertainties 
in the NRQCD matrix elements and large corrections in NLO in $\als$.
The convergence of the perturbative series of the four-fermion matching coefficients 
is indeed often bad (for examples see \cite{decrev}). 
A solution may be provided by the resummation of the 
large contributions in the perturbative series 
coming from bubble-chain diagrams. This analysis has been successfully carried
out in some specific cases in \cite{chen}.

\subsection{Production}
As we have explained in the previous section, colour singlet production and colour singlet 
fragmentation underestimated the data on prompt quarkonium production at Fermilab 
by about an order of magnitude indicating that additional fragmentation contributions 
were missing \cite{prod}. This missing contribution is precisely the gluon fragmentation into 
colour-octet $ ^3S_1$ charm quark pairs. The probability to form a $J/\psi$ particle from a pointlike 
$c\bar{c}$  pair in a colour octet  $ ^3S_1$ state is given by a NRQCD 
 nonperturbative matrix element 
which is suppressed by $v^4$ relative to the leading singlet term but 
 is enhanced by two powers of $\als$ 
 in the short distance part  for producing color-octet quark pairs. When one introduces the leading 
colour-octet contributions, then the data of CDF can be reproduced. 
Still remains a puzzle the behaviour of the polarization at high $p_{\rm t}$ \cite{prod}.

\section{Potential Non Relativistic QCD}

 In NRQCD   the dominant role of the potential as  well as the
 quantum mechanical nature of the problem are not yet maximally
 exploited. A higher  degree of simplification may  be achieved
building  another effective theory for the low
 energy region  of the non-relativistic bound-state, i.e.an  
EFT where only the ultrasoft  degrees of freedom remain dynamical,
 while the rest is integrated out. We integrate out the scale of the momentum transfer $\sim mv$
which is supposed  to be the next relevant scale. Then, two different
situations may exist. In the first one, $mv \gg \Lambda_{\rm QCD}$  and
       the matching from NRQCD to pNRQCD may be performed in
 perturbation theory, expanding in $\alpha_s$.  In the second situation, $mv
  \siml \Lambda_{\rm QCD}$,  the matching has to be nonperturbative,
 i.e. no expansion in $\alpha_s$ is allowed.  Recalling that $r^{-1}
  \sim mv$, these two situations correspond  to systems with inverse
typical radius smaller or  bigger than $\Lambda_{\rm QCD}$, or systems
respectively dominated by the short range or long range (with respect
     to the confinement radius) physics.  

\subsection{$mv \gg \Lambda_{\rm QCD}$}

The effective degrees of
freedom are: $Q\bar{Q}$ states (that can be decomposed into  a singlet
and an octet wave function under color transformations) with energy of
  order of the next relevant  scale, $ \Lambda_{\rm QCD},mv^2$ and
    momentum   ${\bf p}$ of order $mv$,  plus  ultrasoft gluons
	$A_\mu({\bf R},t)$ with energy  and momentum of order
$\lQ,mv^2$. All the  gluon fields are multipole  expanded (i.e.  expanded  in
$r$). The Lagrangian is then  an expansion  in the small quantities  $
 {p/m}$, ${ 1/r  m}$ and    $O(\Lambda_{\rm QCD}, m v^2)\times r$.

The pNRQCD Lagrangian is given 
at the next to leading order in the multipole expansion by \cite{long}:
\begin{eqnarray}
& &\hspace{-4mm}
{L}_{\rm pNRQCD}=
  {\rm Tr} \Biggl\{ {\rm S}^\dagger \left( i\partial_0 - {{\bf p}^2\over m} 
- V_s(r) -\sum_{n=1} {V_s^{(n)}\over m^n} \right) {\rm S} 
\nonumber \\
& & \!\!\!\!\!+ {\rm O}^\dagger \left( iD_0 - {{\bf p}^2\over m} 
- V_o(r) - \sum_{n=1} {V_o^{(n)}\over m^n}  \right) {\rm O} \Biggr\}
\nonumber\\
& & \!\!\!\!\!
 + g V_A ( r) {\rm Tr} \left\{  {\rm O}^\dagger {\bf r} \cdot {\bf E} \,{\rm S}
+ {\rm S}^\dagger {\bf r} \cdot {\bf E} \,{\rm O} \right\} 
\label{pnrqcd0}\\
& &\!\!\!\!\!\!\!
   + g {V_B (r) \over 2} {\rm Tr} \left\{  {\rm O}^\dagger {\bf r} \cdot {\bf E} \, {\rm O} 
+ {\rm O}^\dagger {\rm O} {\bf r} \cdot {\bf E}  \right\} -{1\over 4} F^a_{\mu\nu}
F^{\mu \nu a}.  
\nonumber
\end{eqnarray}
At the leading order in the multipole expansion,
the singlet sector of the Lagrangian gives rise to equations of motion of the 
Schr\"odinger type. The two last lines of (\ref{pnrqcd0})
contain (apart from the Yang-Mills Lagrangian) retardation (or non-potential) effects that 
start at the NLO in the multipole expansion. At this order the non-potential
effects come from the singlet-octet and octet-octet 
interactions mediated by an ultrasoft chromoelectric 
field. 

Recalling that  
 ${ r} \sim 1/mv$ and that the operators count like the next relevant 
scale, $O(mv^2,\lQ)$, to the power of the dimension, it follows that  
each term in  the pNRQCD Lagrangian has a definite power counting.  
From the power 
counting e.g.,  it follows that the interaction of quarks with ultrasoft 
gluons is suppressed in the Lagrangian
 by $v$ ( by $g v$ if $mv^2 \gg \lQ$)   with respect to the LO.

The singlet and octet potentials are well defined objects 
to be calculated in the perturbative matching. In this way a determination
of the singlet $Q\bar{Q}$ potential at three loops leading log has been obtained in 
\cite{logs} and consequently also a determination of $\alpha_V$ which shows how 
this quantity starts to depend on the infrared behaviour  of the theory at three loops.

\subsubsection{Spectrum}

Given the Lagrangian in (\ref{pnrqcd0}) it is possible to calculate the 
quarkonium energy levels at order $m\als^5$ \cite{logs,logss}. At this order  the energy $E_n$
of the level $n$ receives  both from the average value of the potentials
 and  from the singlet-octet interaction (retardation 
effect) a contribution that read
\begin{equation}
E_{n}  =  \langle n\vert V_s(\mu )\vert 
 n \rangle 
-i g^2   \!\! \int_0^\infty \!\!\! dt \,  
 \langle n|  {\bf r} e^{it( H_s - H_o)} {\bf r}  | n \rangle  
\;  \langle {\bf E} (t) \,   {\bf E}  (0)  \rangle(\mu)
\label{en}
\end{equation}
being $H_S$ abd $H_o$ the singlet and octet Hamiltonian respectively.
The nonlocal electric correlator 
is a nonperturbative object dominated by the scale $\lQ$,
$\langle {\bf E} (t) \,   {\bf E}  (0)  \rangle(\mu)\rangle \sim \exp\{-\lQ t\}$.
Thus the integral in (\ref{en}) is the convolution of two exponentials with 
exponent of order  $mv^2$ (for the energy) or  $\lQ$ (for the correlator).
Depending on the relative relation of the two scales three different situations 
take place:
\begin{itemize}
\item{} if $mv^2 \gg \lQ$ then the  correlator  reduces to the local gluon correlator
 and the  second contribution in (\ref{en}) corresponds to the previously mentioned Voloshin-Leutwyler 
 sum rule contribution \cite{sumrul2}.
\item{} if $mv^2 \ll \lQ$ then the energy exponential can be expanded and the 
second contribution in (\ref{en}) corresponds to a short range nonperturbative potential 
corrections \cite{long}.
\item{} if $mv^2 \sim \lQ$ then neither exponentials can be expanded and 
 the nonlocal condensate has to  be used in the energy level calculation \cite{long}.
\end{itemize}
Both the potential model and the sum rule results are contained in  pNRQCD 
as different kinematical limits. What appeared as a puzzle and a problem 
at the origin is now understood as a consequence of the richness of 
the quarkonium dynamics and is appropriately accounted for by the EFT.

The calculation of the quarkonium energy levels at higher orders in 
perturbation theory is relevant to extract the masses of the heavy quarks 
from the $\Upsilon(1S)$, $J/\psi$ and $ttbar$ production cross section
cf. \cite{mass,logss}. 
The perturbative determination of the levels, have been used in \cite{app1,app2}
for  the calculation of the energy levels of some lowest resonances 
of bottomonium, charmonium and $B_c$, after having removed the renormalon
(between the pole mass and the singlet static potential) and under the assumption 
(or to test the assumption)
that the nonperturbative corrections, in the form of nonlocal condensates or 
short range nonperturbative potentials, are  in total  small.

\subsection{$\lQ \sim mv$}
In this case the (nonperturbative) 
matching to pNRQCD has to be done in one single step \cite{M12}.
Under the circumstances that other degrees of freedom 
(like those associated 
with heavy-light meson pair  threshold production and heavy hybrids) 
develop a mass gap of order $\lQ$ 
the quarkonium singlet field $\rm S$ remains as the only dynamical 
degree of freedom in the pNRQCD Lagrangian, which reads \cite{long,M12,sw}
\begin{equation}
\quad  {L}_{\rm pNRQCD}= {\rm Tr} \,\Big\{ {\rm S}^\dagger
   \left(i\partial_0-{{\bf p}^2 \over
 2m}-V_S(r)\right){\rm S}  \Big \}
\end{equation}
In this regime   we recover the quark potential singlet model from
 pNRQCD \cite{M12,sw}.  The final result for the potentials (static and
relativistic corrections) appears factorized in a part  containing the
high energy  dynamics (and calculable in perturbation theory) which is
 inherited from the NRQCD matching coefficients, and a part containing the low energy dynamics
given in terms of Wilson loops and chromo-electric and chromo-magnetic
 insertions in the Wilson loop \cite{M12}.  Such low
 energy contributions can be  calculated  on the lattice \cite{bal}  or
 evaluated in QCD vacuum models \cite{rev,vac}.
The expression obtained for the potential {\it is} the QCD expression, in particular 
all the perturbative contributions to the potential at the hard scale are correctly taken into account.
This solves the problem of consistency with perturbative one-loop calculations
 that was previously encountered in the Wilson loop approach. Moreover, 
further contributions, including a $1/m$ nonperturbative potential,
 appear with respect to the 
Wilson loop original results \cite{wilsonpot}.

\subsubsection{Decays}
The inclusive quarkonium decay width  
 achieve in pNRQCD a factorization with respect to the wave function 
(or its derivatives) calculated in zero
which  is suggestive of the early potential models results:
(cf. eq. (\ref{decpot}))
\be
  \Gamma ({\rm H}\to{\rm LH}) = F(\als,\lQ) \, \cdot \, \vert \psi (0)\vert^2 .
\label{pfac}
\ee
Similar expressions hold for the electromagnetic decays.
However, 
the coefficient $F$ depends here  both on $\als$ and $\lQ$. In particular 
all NRQCD matrix elements, including the octet
ones, can be expressed through pNRQCD as products of  universal 
nonperturbative factors by the squares of the quarkonium wave functions
(or derivatives of it) at the origin. The nonperturbative factors are typically
integral of nonlocal electric or magnetic correlators and thus  
depending on the glue but not on the quarkonium state \cite{sw}. The presence of this 
nonperturbative correlators is indicated by the $\lQ$ dependence of $F$ in
 (\ref{pfac}). Typically $F$ contains both the NRQCD matching coefficients $f$ at the 
hard scale $m$ and the nonperturbative correlators at the low energy scale $\lQ$.
The  nonperturbative correlators, being state independent, are in a
smaller number than the   nonperturbative NRQCD   matrix elements
and thus the predictive power is greatly increased in going from NRQCD to pNRQCD.
   In \cite{sw}  the inclusive decay widths into light hadrons, 
photons and lepton pairs of all $S$-wave and $P$-wave states (under threshold) 
have been calculated up to ${\mathcal O}(mv^3\times (\Lambda_{\rm QCD}^2/m^2,E/m))$ 
and ${\mathcal O}(mv^5)$. A  large reduction in
the number of unknown nonperturbative parameters is achieved and, therefore,
after having fixed the nonperturbative parameters on charmonium decays, 
new model-independent QCD predictions are given for the bottomonium decay
widths \cite{sw}.

Once the methodology to compute the potentials  (real and imaginary contributions) and 
from these the inclusive decays, within a 
$1/m$ expansion in the matching has been developed, the next question to be
addressed
 is  to which extent one can compute the {\it full} potential 
within a $1/m$ expansion in the case $\lQ \gg mv^2$.
It has been shown \cite{sw1} that  new non-analytic terms 
arise due to the three-momentum scale $\sqrt{m\lQ}$. These terms 
can be incorporated into local potentials ($\delta^3 ({\bf r})$ and derivatives of it) 
and scale as half-integer powers of $1/m$. Moreover, 
it is possible to factorize these effects in a model independent 
way and compute them within a systematic expansion in some small parameters \cite{sw1}.

\subsubsection{Production}
Since the power counting of pNRQCD
may be different from the power counting of NRQCD, we expect  that we
 may eventually explain  in this way some of the  difficulties that NRQCD is
 facing in explaining the polarization of the prompt $J/\psi$ data. In particular, if the 
magnetic field turns out to be not suppressed with respect to the electric field operator 
in the power counting, then the spin  flip term is enhanced and the polarization  
may be diluted explaining the behaviour of the data with high $p_t$ \cite{sw,M12,fleming}.

\subsection{Renormalization group improvement and Poincar\'e invariance}

The effective field theory can be used for a very efficient resummation of 
the large logs (typically logs of the ratio of energy and momentum scales)
once a renormalization groups analysis of the EFT has been performed. Such program 
has been successfully performed in pNRQCD \cite{rg}.

\vskip 0.01truecm
Since the EFTs are constructed to be equivalent to QCD, Poincar\'e invariance 
has still to hold. In \cite{poincare} the constraints induced by 
the algebra of the Poincar\'e 
generators on non-relativistic effective field theories have been discussed.
It has been shown that Poincar\'e invariance imposes well defined relations among 
the EFT matching coefficients. The relations have been given both 
for  NRQCD and for pNRQCD.

\section{Conclusions}
The progress in our understanding of non-relativistic effective field theories
makes it possible to move beyond {\em ad hoc} phenomenological models and have 
a unified description of the different heavy-quarkonium observables, 
so that the same quantities determined from a set of data may be used in order to
describe other sets. The old puzzles and problems have been clarified and have 
been understood inside the EFT formulation.
 Predictions based on non-relativistic EFTs 
are conceptually solid, and systematically improvable. 
EFTs put quarkonium on the solid ground of QCD: 
 quarkonium becomes a  privileged window  for precision measurements,
 new physics, confinement mechanism investigations \cite{qwg}.

\section*{Acknowledgments}
We thank  the Organizers for the very nice, interesting and stimulating 
Conference.

\end{document}